\documentclass[
    ,final            
  ,draft            
  ]
  {aipproc}

\layoutstyle{6x9}

\begin{document}

\title{SNe feedback and the formation of elliptical galaxies}


\author{Antonio Pipino}{
  address={Dipartimento di Astronomia, Universit\`a di Trieste}
}

\author{Francesca Matteucci}{
  address={Dipartimento di Astronomia, Universit\`a di Trieste}
}

\begin{abstract}
 The processes governing both the formation and evolution of elliptical galaxies are
discussed by means of a new multi-zone photo-chemical evolution model for
elliptical galaxies, taking into account detailed nucleosynthetic
yields, feedback from supernovae, Pop III stars and an initial infall
episode. 

By comparing model predictions with observations, we derive a
picture of galaxy formation in which the higher is the mass of the
galaxy, the shorter are the infall and the star formation
timescales.  
In particular, by means of our model,
we are able to reproduce the overabundance of Mg relative to Fe,
observed in the nuclei of bright ellipticals, and its increase with
galactic mass.  

This is a clear sign of an anti-hierarchical formation
process. Therefore, in this scenario, the most massive objects are
older than the less massive ones, in the sense that larger galaxies
stop forming stars at earlier times. 

Each galaxy is created
outside-in, i.e. the outermost regions accrete gas, form stars and
develop a galactic wind very quickly, compared to the central core in
which the star formation can last up to $\sim 1.3$ Gyr.  This
finding will be discussed at the light of recent observations of the
galaxy NGC 4697 which clearly show a strong radial gradient in
the mean stellar [$<Mg/Fe>$] ratio.    

The role of galactic winds in the IGM/ICM enrichment will
also be discussed.

\end{abstract}

\maketitle


\section{Introduction}
Any model of galaxy evolution presented so far had to overcome the strong challenge
represented by the observational fact that elliptical galaxies show a remarkable 
uniformity in their photometric and chemical properties.
Metallicity gradients are characteristic of the stellar populations
inside elliptical galaxies. Evidences come from the increase of
line-strength indices and the reddening of the colours towards the centre of the galaxies
(for details and references see Pipino \& Matteucci 2004, PM04).  The study of such
gradients provide insights into the mechanism of galaxy
formation, particularly on the duration of the chemical enrichment
process at each radius.  Metallicity indices, in fact, contain
information on the chemical composition and the
age of the single stellar populations (SSP) inhabiting a given
galactic zone.  In particular, by comparing indices related mainly to
Mg to others representative of the Fe abundance, it is possible to derive the [Mg/Fe] abundance ratio,
which is a very strong constraint for the formation timescale of a
galaxy.  In fact, the common interpretation of the 
$\alpha$-element (O, Mg, Ca,
Si) overabundance relative to Fe, and its decrease with increasing
metallicity in the solar neighbourhood
is due to the different origin of these elements
(time-delay model, Matteucci $\&$ Greggio, 1986), being the former
promptly released by type II supernovae (SNII) and the latter mainly
produced by type Ia supernovae (SNIa) on longer timescales. 
The time-delay model applies also to other objects 
and the [$\alpha$/Fe] versus [Fe/H] relation depends strongly on the star formation history.
For a very short and intense star burst, the [$\alpha$/Fe] ratios
decrease at higher metallicity than in the solar vicinity.
PM04 showed that a galaxy formation process in which the most massive
objects form faster and more efficiently than the less massive ones
can explain the photo-chemical properties of ellipticals, in 
particular the increase of [Mg/Fe] ratio in stars with galactic mass 
(see PM04). Moreover,
from an extended analysis of metallicity and colour gradients, Pipino, Matteucci \&
Chiappini (2006, PMC) suggest that a single galaxy should form outside-in, namely the
outermost regions form earlier and faster with respect to the central
parts.  A natural consequence of this
model and of the time-delay between the production of Fe and that of Mg
is that the mean [Mg/Fe] abundance ratio in the stars should
increase with radius.

\section{The model}

The adopted chemical evolution model is based on that
presented by PM04. In this particular case we consider our model galaxies
as a multi-zone extending out to 10 effective radii, with instantaneous mixing of gas. Moreover
we take explicitly into account a possible mass flow due to the galactic wind
and a possible secondary episode of gas accretion
in order to model late time gas accretion and/or interactions
with the environment. The chemical code features a new
self-consistent energy treatment which supersedes the previous
one adopted by PM04 (see Pipino et al., 2005).
Particular care is dedicated
to a detailed calculations of Type Ia and II SN rates.
The minimum SN efficiency required to develop a galactic wind is 10\%.

\section{Results and conclusions}

We find that 
SF and infall timescales decreasing with galactic mass are needed to explain the optical properties
of elliptical galaxies (PM04). 
At the same time we reproduce the $L_X - L_B$ relation in the ISM of bright ellipticals (Pipino et al., 2005). 
Our best model satisfies the main constraint represented by the observed [$\rm <\alpha/Fe>_V]-\sigma$
relation (see Fig. 1, left panel).

\begin{figure}
\includegraphics[width=6cm]{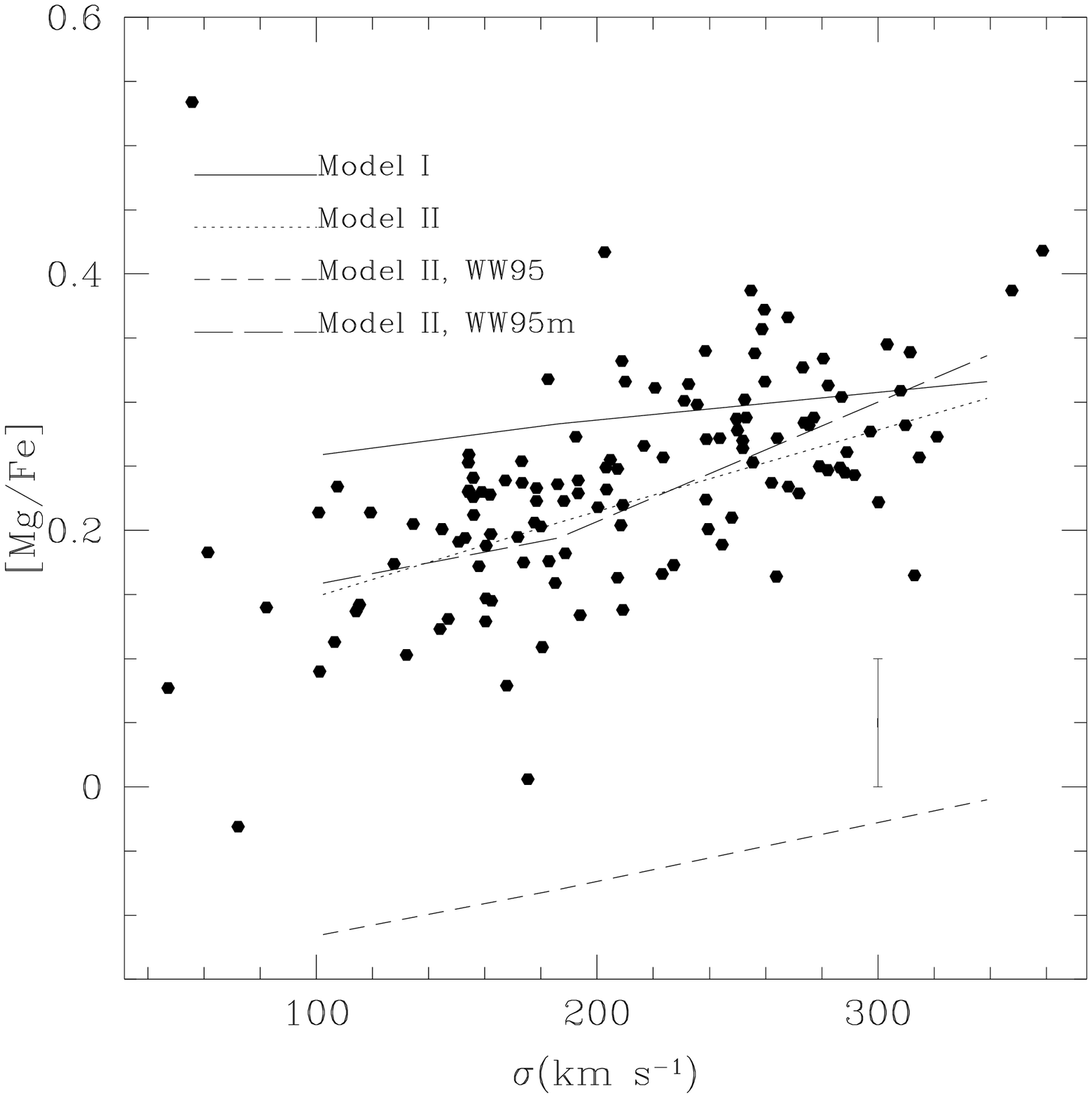}
\includegraphics[width=6cm]{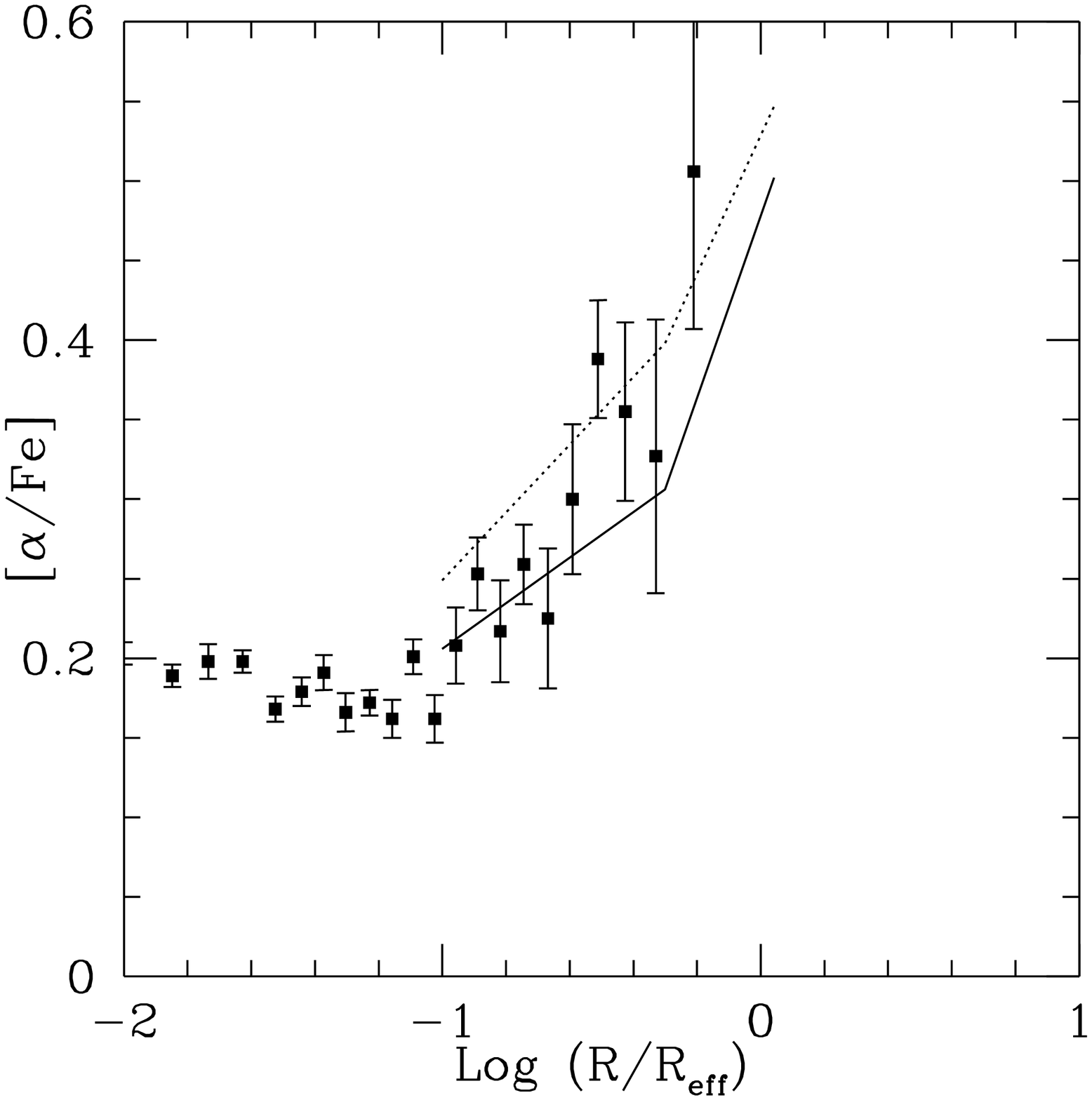}
\caption{\emph{Left:} [Mg/Fe] as a function of galactic velocity dispersion predicted by Model I (solid)
and II (dotted) compared to the
data from Thomas, Maraston, $\&$ Bender (2002). 
The typical error is shown in the bottom-right corner. For comparison we show the
theoretical curves obtained with the same input parameters of Model II, but different yields (see text).
WW95: yields by Woosley $\&$ Weaver (1995). WW95m: modified yields by Woosley $\&$ Weaver (1995), see PM04.
\emph{Right:} PM04's model IIb predictions for the mean mass-weighted [$\rm <\alpha/Fe>$] (solid) and 
luminosity-weighted [$\rm <\alpha/Fe>_V$] (dotted) abundance ratios
in stars as a function of radius compared to the [$\alpha$/Fe] derived for the galaxy NGC 4697 (Mendez et al. 2005, full
squares).}
\end{figure}

In Fig. 1 (right panel) we show the predicted radial trends of both the [$\rm <Mg/Fe>$] (solid line) and [$\rm <Mg/Fe>_V$] (dotted line) 
abundance ratios versus the
\emph{observed} one in NGC 4697.  The latter is obtained by Mendez et al. (2005) by converting
the line-strength indices into abundances. The agreement is remarkable,
especially because we did not tune the input parameters (i.e. 
radius, mass) to exactly match NGC 4697. 
The observed increase of
[Mg/Fe] with radius confirm
PM04's model predictions, namely an outside-in formation process
in which the central part of the galaxy form stars for
a longer period compared to the most external regions.
This can be explained in terms of galactic winds developing
earlier where the local potential well is shallower.

By comparing the radial trend of [$\rm <Z/H>$] with the 
\emph{observed} one,
we notice a discrepancy which is due to the fact that a
CSP behaves in a different way with respect to a SSP.
In particular the predicted gradient of [$\rm <Z/H>$] is flatter
than the observed one at large radii.
Therefore, this should be taken into account when 
estimates for the metallicity of a galaxy are derived
from the simple comparison between the observed line-strength index
and the prediction for a SSP, a method currently adopted in the literature (see PMC).

The new energy formalism implemented in the chemical evolution code allows us to follow in a 
more detailed way the evolution of mass and energy flow
into the ICM with respect to previous works.
The predicted amount of Fe ejected by ellipticals into the ICM match the observations, 
and new data on the [$\alpha$/Fe] ratios are in better agreement with our results (Pipino et al., 2005).
Therefore, we confirm that SNe Ia
are fundamental in providing energy and iron to the ICM.


\begin{theacknowledgments}
The work was supported by MIUR under COFIN03 prot. 2003028039.
A.P. thanks the Organizers for having provided financial support for attending the conference.
\end{theacknowledgments}



\bibliographystyle{aipprocl} 

\bibliography{sample}

 {\typeout{}
  \typeout{******************************************}
  \typeout{** Please run "bibtex \jobname" to optain}
  \typeout{** the bibliography and then re-run LaTeX}
  \typeout{** twice to fix the references!}
  \typeout{******************************************}
  \typeout{}
 }

\end{document}